\documentclass[twocolumn,prl,showpacs,superscriptaddress,nofootinbib]{revtex4-2}
\usepackage[latin9]{inputenc}
\usepackage{amsmath}
\usepackage{amssymb}
\usepackage{amsthm}
\usepackage{graphicx}
\usepackage{color}

\theoremstyle{plain}
\newtheorem*{theorem*}{Theorem}

\makeatletter

\@ifundefined{textcolor}{}
{%
 \definecolor{BLACK}{gray}{0}
 \definecolor{WHITE}{gray}{1}
 \definecolor{RED}{rgb}{1,0,0}
 \definecolor{GREEN}{rgb}{0,1,0}
 \definecolor{BLUE}{rgb}{0,0,1}
 \definecolor{CYAN}{cmyk}{1,0,0,0}
 \definecolor{MAGENTA}{cmyk}{0,1,0,0}
 \definecolor{YELLOW}{cmyk}{0,0,1,0}
}

\usepackage{color}
\usepackage{amsfonts}
\usepackage{graphics}
\usepackage{epsfig}\usepackage{amsthm}

\def\identity{\leavevmode\hbox{\small1\kern-3.8pt\normalsize1}}

\renewcommand{\epsilon}{\varepsilon}

\makeatother

\begin{document}

\title{Room Temperature Light-Mediated Long-Range Coupling of Excitons in Perovskites}

\author{Tanjung Krisnanda}
\affiliation{School of Physical and Mathematical Sciences, Nanyang Technological University, 637371 Singapore, Singapore}

\author{Qiannan Zhang}
\affiliation{School of Physical and Mathematical Sciences, Nanyang Technological University, 637371 Singapore, Singapore}

\author{Kevin Dini}
\affiliation{School of Physical and Mathematical Sciences, Nanyang Technological University, 637371 Singapore, Singapore}

\author{David Giovanni}
\affiliation{School of Physical and Mathematical Sciences, Nanyang Technological University, 637371 Singapore, Singapore}

\author{Timothy C. H. Liew}
\affiliation{School of Physical and Mathematical Sciences, Nanyang Technological University, 637371 Singapore, Singapore}
\affiliation{MajuLab, International Joint Research Unit UMI 3654, CNRS, Universit\'{e} C\^{o}te d'Azur, Sorbonne Universit\'{e}, National University of Singapore, Nanyang Technological University, Singapore}

\author{Tze Chien Sum}
\affiliation{School of Physical and Mathematical Sciences, Nanyang Technological University, 637371 Singapore, Singapore}

\begin{abstract}
Perovskites have been the focus of attention due to their multitude of outstanding optoelectronic properties and structural versatility. 
Two-dimensional halide perovskite such as $(\mbox{C}_6\mbox{H}_5\mbox{C}_2\mbox{H}_4\mbox{N}\mbox{H}_3)_2\mbox{Pb}\mbox{I}_4$, or simply PEPI, forms natural multiple quantum wells with enhanced light-matter interactions, making them attractive systems for further investigation. 
This work reports tunable splitting of exciton modes in PEPI resulting from strong light-matter interactions, manifested as multiple dips (modes) in the reflection spectra.
While the origin of the redder mode is well understood, that for the bluer dip at room temperature is still lacking. 
Here, it is revealed that the presence of the multiple modes originates from an indirect coupling between excitons in different quantum wells. 
The long-range characteristic of the mediated coupling between excitons in distant quantum wells is also demonstrated in a structure design along with its tunability.
Moreover, a device architecture involving an end silver layer enhances the two excitonic modes and provides further tunability. 
Importantly, this work will motivate the possibility of coupling of the excitonic modes with a confined light mode in a microcavity to produce multiple exciton-polariton modes.

\begin{flushleft}

\noindent {\bf Correspondence}: T.K. (tanjung.krisnanda@ntu.edu.sg); T.C.H.L.(TimothyLiew@ntu.edu.sg); or T.C.S. (tzechien@ntu.edu.sg)\\

\noindent T.K., Q.Z., K.D., and D.G. contributed equally to this work.\\

\end{flushleft}
\end{abstract}

\maketitle

\section{Introduction}
Solution-processed lead halide perovskites provide an alternative for photovoltaic cells.\textsuperscript{\cite{kojima2009organometal,green2014emergence}}
They offer a cheap and efficient option, making them competitive candidates and resulting in a rapidly evolving technology in the field.\textsuperscript{\cite{gratzel2017rise,correa2017promises}}
Owing to their excellent optoelectronic properties and structural versatility, these materials have also been proven useful for other applications, for example: LEDs,\textsuperscript{\cite{wang2016perovskite, zhao2018high, shang2019highly, yang2020efficient}} ferroelectrics,\textsuperscript{\cite{zhang2019tunable}} spintronics as shown by the observation of Rashba splitting\textsuperscript{\cite{zhai2017giant, yin2018layer, todd2019detection, chen2018impact}} as well as spin manipulation,\textsuperscript{\cite{giovanni2016tunable, giovanni2019ultrafast, lu2019spin}} photodetectors,\textsuperscript{\cite{shi2019leap, georgiadou2019high, leung2018self}} and memristors.\textsuperscript{\cite{zhao2019memristors, solanki2019interfacial}}
Lead halide perovskites also offer multidimensional platforms for tuning their optoelectronic properties.
In particular, 2D perovskites offer much stronger excitonic properties and moisture stability as compared to their 3D counterparts.
Due to quantum and dielectric confinement of their natural multi quantum well (MQW) structure, they become hotbeds of exotic photophysics, for instance: strong light-matter interaction in the form of exciton-polaritons,\textsuperscript{\cite{wang2018room,fieramosca2019two}} the optical Stark effect,\textsuperscript{\cite{giovanni2016tunable}} coherent exciton-phonon interactions,\textsuperscript{\cite{davidpepi2018}} the giant Rashba effect,\textsuperscript{\cite{zhai2017giant}} etc.

\vspace{0.3cm}
Herein, we focus on archetypal 2D perovskite $(\mbox{C}_6\mbox{H}_5\mbox{C}_2\mbox{H}_4\mbox{N}\mbox{H}_3)_2\mbox{Pb}\mbox{I}_4$ or $(\mbox{PEA})_2\mbox{Pb}\mbox{I}_4$ (henceforward referred to as PEPI), which shows a high exciton binding energy of $\sim$200 meV.\textsuperscript{\cite{zhai2017giant}}
Motivated by its strong excitonic properties and exotic self-assembled MQW structure, we study the former further for different PEPI designs and demonstrate a tunable excitonic mode splitting, revealed by our steady state reflectivity spectra measurements.
By utilising the transfer matrix formalism, we unveil the origin of such splitting, i.e., it is a consequence of long-range interactions between excitons in different quantum wells that are mediated by light.
Moreover, in a structure design composed of two thin PEPI layers separated by a Polymethylmethacrylate (PMMA) barrier, we show how excitons in distant quantum wells are coupled via light that is propagating between the PEPI layers, and that the indirect interactions can be long-range and tuned by varying the thickness of the barrier.
In a more complex design, which employs PMMA barriers and an end silver layer, we demonstrate the enhancement of the absorption strength of the excitonic modes and also their tunability.

\vspace{0.3cm}
Our results also imply an alternative for constructing exciton-polaritons in microcavities.
In particular, by placing our PEPI designs in a microcavity one can in principle couple the tunable excitonic modes with a well defined cavity light mode.\textsuperscript{\cite{strongexpho1,strongexpho2,strongexpho3,han2020transition}}
It has been shown previously, particularly in inorganic semiconductors, that by using either two exciton\textsuperscript{\cite{cristofolini2012coupling}} or multiple photon modes,\textsuperscript{\cite{diederichs2006parametric,duan2013polariton,xie2012room}} one can achieve multiple exciton-polariton modes. 
This is particularly interesting as with multiple coupled exciton-polariton modes, one can realize parametric oscillation.\textsuperscript{\cite{diederichs2006parametric,xie2012room}}
Theoretically, we also expect this geometry to be a key for the generation of THz radiation\textsuperscript{\cite{kyriienko2013superradiant}} or the enhancement of nonlinearity at the quantum level.\textsuperscript{\cite{kyriienko2014triggered}}

\begin{figure*}[t]
\centering
\includegraphics[width=0.8\textwidth]{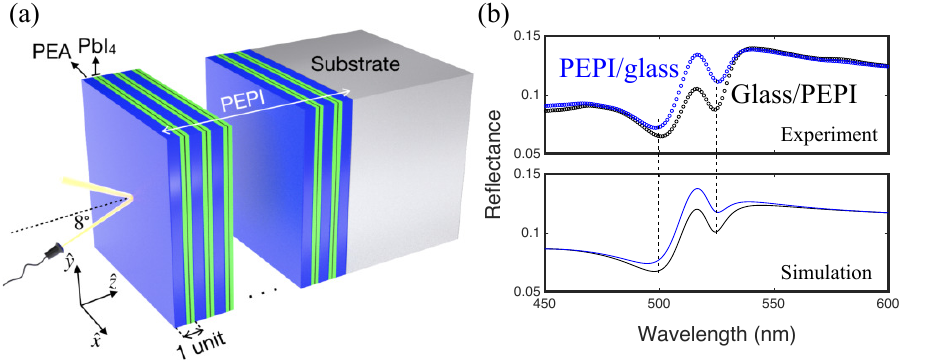}
\caption{
(a) Setup and model of two-dimensional PEPI perovskite. The PEPI is treated as a stack of multiple units, each comprises a PEA and an active layer $\mbox{Pb}\mbox{I}_4$. The active layers contain excitons that can be excited with a resonance frequency $\omega_0$. (b) Reflectivity spectra of PEPI/glass and its reverse configuration. The blue (PEPI/glass) and black (glass/PEPI) circles represent experimental data, and simulations are indicated by the corresponding solid curves. We note that the two notable dips (excitonic modes) are reproduced by simulations.}
\label{FIG_setup}
\end{figure*}

\section{Model formalism}

Here we construct an effective transfer matrix model for the PEPI thin film, which consists of multiple active layers of $\mbox{PbI}_4$ (quantum wells) separated by dielectric barriers (PEA).
In what follows, we study 2D structures, as exemplified in {\bf Figure~\ref{FIG_setup}}(a), where an incident light penetrates the different composing layers.
We will only consider transverse electric (TE) polarized incident light, which is sufficient \textcolor{black}{as we considered small angle of incidence in our experiments.}
The corresponding relevant electric and magnetic field are denoted by $E_x$ and $H_y$, respectively.

\vspace{0.3cm}
Let us start with a 2D dielectric layer, which we shall refer to as layer $M$, having refractive index $n_M(\omega)$ and thickness $L_M$.
The inplane electric and magnetic fields at both ends can be related by the following equation:
$[E_x^{\prime} \:H_y^{\prime}]^T={\bf T}_M\:[E_x \:H_y]^T$ \textcolor{black}{($T$ denotes transposition)},
where the $\hat x\hat y$ plane is parallel to the 2D layer (see~Figure~\ref{FIG_setup}(a)).
Assuming spatial homogeneity of the 2D layer, the fields are uniform in the $\hat x$ and $\hat y$ directions, while they are allowed to vary in the orthogonal $\hat z$ direction (i.e., the material growth direction).
The prime ($^\prime$) indicates the fields right after the light passes through the layer.
From Maxwell's equations under TE boundary conditions, one can express the transfer matrix ${\bf T}_M$ across layer $M$ as\textcolor{black}{\textsuperscript{\cite{Vladimirova_96,kavokin2003cavity,Kavokinbook2017}}}:
\begin{equation}\label{EQ_NTmatrix}
{\bf T}_M=\left[ \begin{array}{cc} \cos(\Delta)&\frac{i}{q_M}\sin(\Delta)\\ iq_M \sin(\Delta)& \cos(\Delta) \end{array}\right],
\end{equation}
where $\Delta=n_M(\omega)k_0L_M\cos(\theta_M)$ and $q_M=n_M(\omega)k_0\cos(\theta_M)/\omega \mu_0$ is the inverse impedance with $k_0=\omega/c$ and $\theta_M$ the propagation angle (with respect to the $\hat z$ axis) in the layer, which one relates to the angle of incidence through the Snell-Descarte law as $\theta_M=\arcsin \left(n_{\scriptsize \mbox{L}}\sin(\theta)/n_{M}(\omega)\right)$.
Here $n_{\scriptsize \mbox{L}}$ denotes the refractive index of the material in which the light is incident, e.g., air, and $\theta$ is the angle of incidence of light.
We model our material system as a stack of multiple PEPI units (see Figure~\ref{FIG_setup}(a)).
The transfer matrix in Equation~(\ref{EQ_NTmatrix}) was used in simulations to describe the PEA, and also for silver and PMMA layers in more complex structures considered later.
\textcolor{black}{Note that we treat the dimension of the fields ($E_x$ and $H_y$) in SI units, and therefore the off-diagonal elements of the transfer matrix are not dimensionless.}

\vspace{0.3cm}
A quantum well (QW) layer reacts differently in the presence of light due to the presence of optically active excitons.
For an infinitely thin quantum well, the TE mode transfer matrix reads\textcolor{black}{\textsuperscript{\cite{vladimirova1998exciton,Kavokinbook2017}}}:
\begin{equation}\label{EQ_QWTmatrix}
{\bf T}_{\scriptsize \mbox{QW}}=\left[ \begin{array}{cc} 1&0\\ \frac{2q_{\scriptsize \mbox{QW}}r_{\scriptsize \mbox{QW}}}{1+r_{\scriptsize \mbox{QW}}}& 1 \end{array}\right],
\end{equation}
where $q_{\scriptsize \mbox{QW}}=\textcolor{black}{n_{\scriptsize \mbox{off}}}k_0\cos(\theta_{\scriptsize \mbox{QW}})/\omega \mu_0$ with \textcolor{black}{$n_{\scriptsize \mbox{off}}$} as the \textcolor{black}{off-resonance} refractive index of the quantum well layer and $\theta_{\scriptsize \mbox{QW}}$ is obtained from the Snell-Descarte law.
\textcolor{black}{We take $n_{\scriptsize \mbox{off}}$ as constant, while the energy dependence near resonance is taken into account in $r_{\scriptsize \mbox{QW}}$ as follows.} 
Considering the Helmoltz equation with a source term including the effect of the excitons in the QW layer, one can obtain the reflection coefficient of the QW\textsuperscript{\cite{vladimirova1998exciton}}:
$r_{\scriptsize \mbox{QW}}=i\Gamma_0/(\omega_0-\omega-i(\Gamma_0+\gamma))$,
where $\Gamma_0$ is the excitonic radiative broadening, $\gamma$ is the homogeneous broadening, and $\hbar \omega_0$ is the resonance energy of the excitons.
\textcolor{black}{The modelling of a quantum well layer with transfer matrix, as in Equation (\ref{EQ_QWTmatrix}), has been routinely used in heterostructures.\textsuperscript{\cite{kavokin2003cavity,Kavokinbook2017}}} 
Although the above formula is for an infinitely thin quantum well, it is a good approximation for our active $\mbox{PbI}_4$ layers as their thickness is smaller than the exciton Bohr radius, such that the account of correlated exciton motion with light in the growth direction is negligible.
The main effect of the quantum well thickness is the contribution to the optical path length.
This can be conveniently accounted for by modelling a single PbI$_4$ as a half PbI$_4$ dielectric layer, using Equation~(\ref{EQ_NTmatrix}), followed by an infinitely thin quantum well layer and another half of PbI$_4$ dielectric layer.

\vspace{0.3cm}
Figure~\ref{FIG_setup}(a) illustrates our model configuration. For one unit of the PEPI layer containing a PEA and an active layer, the transfer matrix of our model reads ${\bf T}_{\scriptsize \mbox{1u}}=({\bf T}_{\scriptsize \mbox{PbI}_4/2}{\bf T}_{\scriptsize \mbox{QW}}{\bf T}_{\scriptsize \mbox{PbI}_4/2}){\bf T}_{\scriptsize \mbox{PEA}}$, where ${\bf T}_{\scriptsize \mbox{PEA}}$ and ${\bf T}_{\scriptsize \mbox{PbI}_4/2}$ denotes the normal-layer transfer matrix for PEA and half-width ${\mbox{PbI}_4}$, respectively.
The complete PEPI layer with $N$ units then has an effective transfer matrix
\begin{equation}\label{EQ_T_PEPI}
{\bf T}_{\scriptsize \mbox{PEPI}}={\bf T}_{\scriptsize \mbox{PEA}}({\bf T}_{\scriptsize \mbox{1u}})^N.
\end{equation}
The thickness of the PEPI and the number of units are related through $L_{\scriptsize \mbox{PEPI}}=N(L_{\scriptsize \mbox{PEA}}+L_{\scriptsize \mbox{PbI}_4})+L_{\scriptsize \mbox{PEA}}$.
From a given effective transfer matrix of a 2D structure situated between a left medium with refractive index $n_{\scriptsize \mbox{L}}$ and a right medium with $n_{\scriptsize \mbox{R}}$, one can construct the corresponding reflectance (see Appendix~\ref{M_reflectance}).

\section{Results}

We compare our model simulations with the actual reflectance spectra. 
The contribution from the normal layers (i.e., glass, silver, and PMMA) and their refractive index dispersion are considered - see Appendix~\ref{M_refractiveindex}, for details.
Experimental information including sample fabrications and measurements are presented in Experimental Section.

\subsection{Bare PEPI thin film}

\begin{figure*}[t]
\centering
\includegraphics[width=0.9\textwidth]{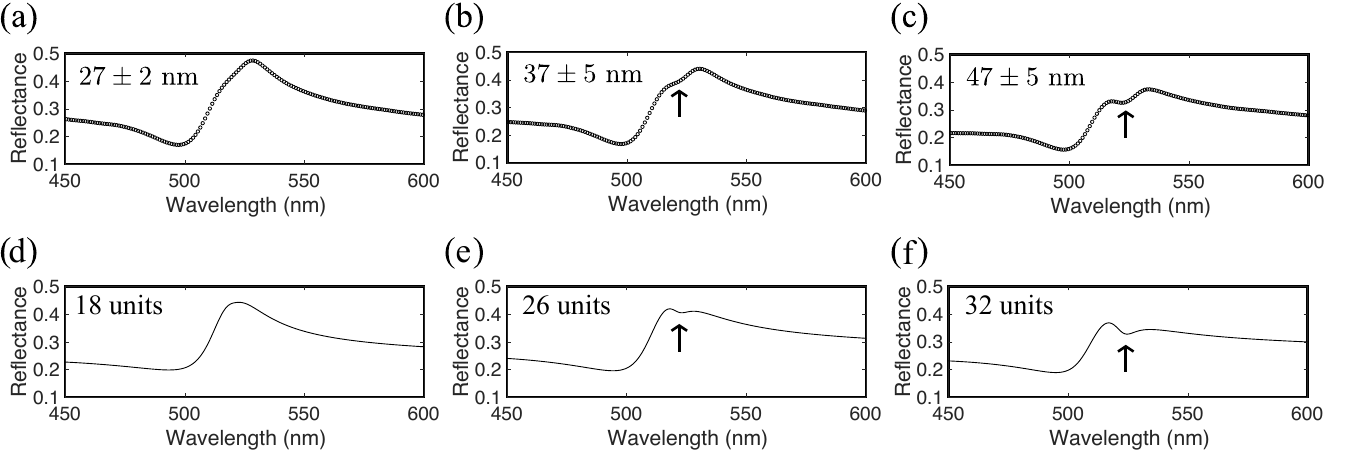}
\caption{Reflectivity spectra of PEPI with different thickness on glass substrate.
Experimental data for $L_{\scriptsize \mbox{PEPI}}=27\pm2$~nm, $37\pm5$~nm, and $47\pm5$~nm are presented in panels (a), (b), and (c), respectively.
The simulations are plotted in panels (d)-(f) with the number of units corresponding to the thickness of the PEPI in (a)-(c), respectively.
We note that experiments and simulations coincide on the formation and the position of the second excitonic mode.
}
\label{FIG_pepiDT}
\end{figure*}

\begin{figure*}[t]
\centering
\includegraphics[width=0.9\textwidth]{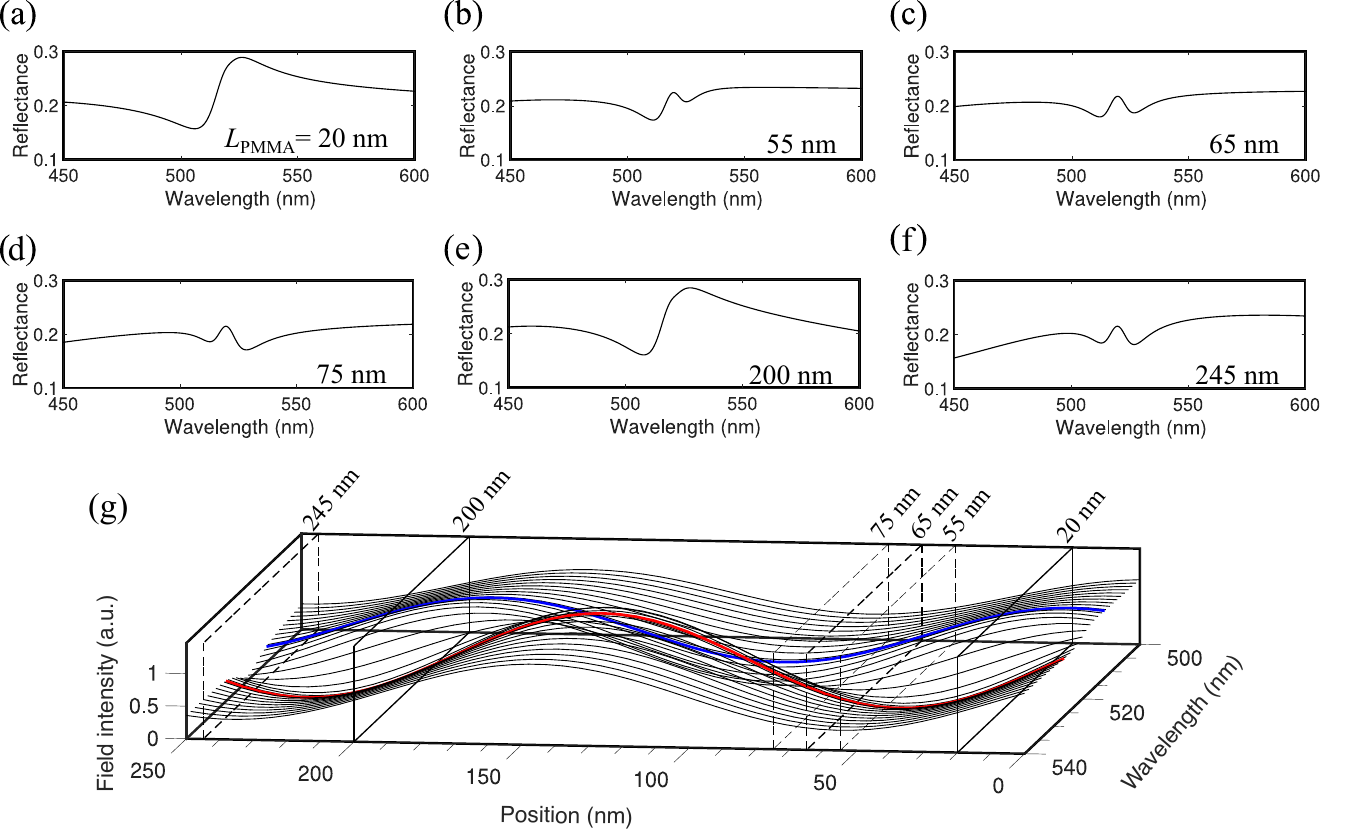}
\caption{Reflectivity spectra and field intensity in PEPI/PMMA/PEPI structure.
Panels (a)-({\color{black}{f}}) show the reflectance for varying values of $L_{\scriptsize \mbox{PMMA}}$, which is indicated at the bottom right of each graph.
{\color{black}{One mode is obtained in panels (a) and (e), whereas two modes are observed in panels (b)-(d) and (f).}}
Panel ({\color{black}{g}}) shows the field intensity ($|E_x|^2$) {\color{black}{in the PMMA barrier after passing the first PEPI layer at $0$~nm.
The solid (one mode) and dashed (two modes) boxes correspond to the position of the second PEPI layer, i.e., the situation in panels (a)-(f).
T}}he axis Position refers to the position in the barrier.
The red {\color{black}{and blue}} curves indicate the two energy modes (position of the two dips).
As a result of passing through the first PEPI at position $0$~nm, the field intensity changes differently for these two modes, through the barrier and when they are reaching the position of the second PEPI layer.
{\color{black}{This mechanism is responsible for the one-two mode switching.}}
}
\label{FIG_pepitoymodel}
\end{figure*}

We proceed to simulate the reflectance by using the transfer matrix model of PEPI in Equation~(\ref{EQ_T_PEPI}).
For our simulations, we use PEA layers with width $L_{\scriptsize \mbox{PEA}}=1$~nm and PbI$_4$ layers with $L_{\scriptsize \mbox{PbI}_4}=0.63$~nm.\textsuperscript{\cite{davidpepi2018}} 
\textcolor{black}{The refractive index of PEA layers is treated to be non-dispersive $n_{\scriptsize \mbox{PEA}}=1.8$, and the off-resonance value for the PbI$_4$ is taken as $n_{\scriptsize \mbox{off}}=2.5$.\textsuperscript{\cite{hong1992dielectric}}}
\textcolor{black}{We note that the refractive index of PbI$_4$ is dispersive near resonance, and in our formalism, this dependence is treated explicitly in $r_{\scriptsize \mbox{QW}}$.}
\textcolor{black}{See also a recent work\textsuperscript{\cite{decrescent2019optical}} indicating the dispersive (effective total) refractive index of PbI$_4$.}
For the quantum well transfer matrix, i.e., ${\bf T}_{\scriptsize \mbox{QW}}$ in Equation~(\ref{EQ_QWTmatrix}), we set the $\omega_0$, $\Gamma_0$, $\gamma$, and the number of units $N$ as fitting parameters.
Note that in our fitting, we also take into account the non-ideality of the detector and the background signal, which were present in experiments (see Appendix~\ref{M_fitting}).

\vspace{0.3cm}
The results are presented in {Figure~\ref{FIG_setup}}(b), where we compare our simulations with the experimental reflectance spectra of PEPI sample spin-coated on a glass substrate, with two orientations: PEPI/glass (blue circles) and its reverse orientation glass/PEPI (black circles).
The parameters used to simulate the spectra are summarized in {\bf Table \ref{tableFitting}}.
Interestingly, we observed two well-separated exciton modes shown by two dips around $\sim500$~nm and $\sim525$~nm, implying an excitonic mode splitting.
Such features are well reproduced by our simulation, thus validating the accuracy of our model.
Our analysis reveals that such excitonic mode splitting is caused by light-mediated coupling \textcolor{black}{(indirect coupling mechanism)} of excitons in different PEPI active layers. 
\textcolor{black}{We note that there is no direct exciton-exciton coupling mechanism (Coulomb coupling) used in our formalism.
The PEA layers are too thick, such that the direct coupling is assumed insignificant.}

\begin{table}[!h]
\begin{center}
\caption{Summary of parameters of the PEPI thin film reflection spectra.
}
\label{tableFitting}
\smallskip
\begin{tabular}{|c||c|c|c|c|c|c|}
\hline
Parameter&$n_{\scriptsize \mbox{PEA}}$&\textcolor{black}{$n_{\scriptsize \mbox{off}}$}&$\hbar\omega_0$&$\hbar\Gamma_0$&$\hbar\gamma$&$N$\\
\hline
Value& 1.8&2.5& 2.387 [eV]&1.27 [meV]&22 [meV]&35 \\
\hline
\end{tabular}
\end{center}
\end{table}

\vspace{0.3cm}
To delve deeper into the physics, we study the effect of PEPI layer thickness on the exciton mode splitting, specifically, on the emergence of the second mode, i.e., the dip around $\sim525$~nm.
Our results show that a thicker PEPI layer, i.e., with more composing units, exhibits more apparent excitonic mode splitting.
The reflectivity spectra for PEPI layers with different thickness are presented in {\bf Figure~\ref{FIG_pepiDT}} panels (a)-(c) for experimental data and panels (d)-(f) for the corresponding simulations, respectively.
It can be seen from the simulations, corroborated by the experimental results, that the second mode (indicated by the arrow) is forming as the thickness of the PEPI layer is increased.
Our results also show that, for a thin PEPI layer, such as the one having less than 18 units or $L_{\scriptsize \mbox{PEPI}}<29$~nm, only one excitonic mode is observed.
This is due to the {\color{black}{indirect}} coupling between excitons, leading to the mode splitting, being overcome by the linewidth of one PEPI unit.

\begin{figure*}[t]
\centering
\includegraphics[width=0.9\textwidth]{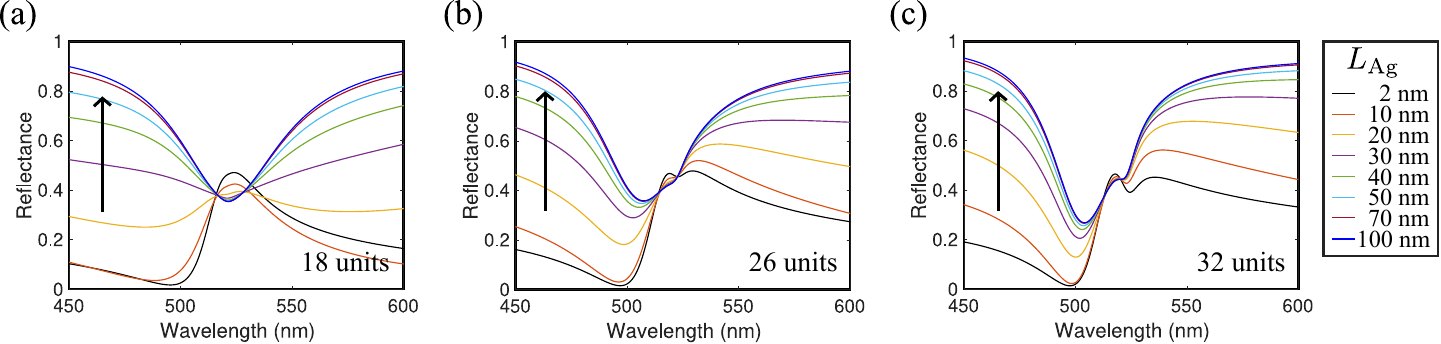}
\caption{Reflectivity spectra of PMMA/PEPI/PMMA/Ag/glass substrate for different thickness of the silver layer. (a) 18 units, (b) 26 units, and (c) 32 units composing the PEPI thin film. The thickness of the Ag layer is increased in the direction as indicated by the arrows.
}
\label{FIG_pepiID}
\end{figure*}

\vspace{0.3cm}
Another factor that is crucial for the splitting is the position of the active layers.
We demonstrate this more closely with a model where we utilize a barrier, i.e., a PMMA layer, sandwiched between two thin PEPI layers.
For our simulations, we used the same parameters ($n_{\scriptsize \mbox{PEA}},\textcolor{black}{n_{\scriptsize \mbox{off}}},\omega_0,\Gamma_0,$ and $\gamma$) as in Table.~\ref{tableFitting}.
We set each PEPI layer to be thin (i.e., having 5 units) such that it can only exhibit one excitonic mode by itself.
The results are shown in {\bf Figure~\ref{FIG_pepitoymodel}} panels (a)-({\color{black}{f}}). 
In such a system where the two PEPI layers {\color{black}{are separated further apart}}, the presence of mode splitting can still be clearly seen.
{\color{black}{This splitting is due to indirect (mediated) interactions via light propagating between the two PEPI layers.}}
Moreover, the relative strength of such mode splitting can be tuned by varying the PMMA thickness~($L_{\scriptsize \mbox{PMMA}}$).

\vspace{0.3cm}
By further increasing $L_{\scriptsize \mbox{PMMA}}$, one can continuously switch between {\color{black}{having}} one or two exciton modes.
This {\color{black}{one-two mode}} switch, which oscillates with respect to the barrier thickness {\color{black}{(see Figure 3 panels (a)-(f))}}, still applies for thicker PMMA layer, and therefore justifies the long-range characteristic of the indirect coupling between excitons.
The mechanism of the indirect coupling can be further seen from Figure~\ref{FIG_pepitoymodel}({\color{black}{g}}).
Here, we plot how the field intensity ($|E_x|^2$) of the probing light changes {\color{black}{(after passing the first PEPI layer at position $0$~nm) as it propagates through the barrier.
The boxes indicate the position of the second PEPI layer, corresponding to the situation in panels (a)-(f).
We have indicated with solid (dashed) boxes the case resulting in one (two) exciton modes.}}
It is apparent that from position $0$~nm the field intensity oscillates differently across the barrier, due to the presence of the first PEPI.
{\color{black}{We have also indicated the two energy modes with solid red and blue curves.
Keeping in mind the oscillation of the field intensity in the barrier, it is clear}} that the second thin PEPI would either receive a different field information or a similar one depending on the thickness of the barrier.
This is in line with the one-two mode switching behaviour described previously.
{\color{black}{In particular, one mode is achieved with $L_{\scriptsize \mbox{PMMA}}=20$~nm and $200$~nm when the higher mode field intensity (blue curve) is much higher than the lower mode (red curve).
On the other hand, two modes are achieved with $L_{\scriptsize \mbox{PMMA}}=65$~nm and $245$~nm when both modes have comparable field intensity.
Furthermore, by varying the PMMA thickness around $65$~nm (i.e., $55$~nm and $75$~nm), one can tune the relative strength of the two modes.}}

\begin{figure*}[t]
\centering
\includegraphics[width=0.9\textwidth]{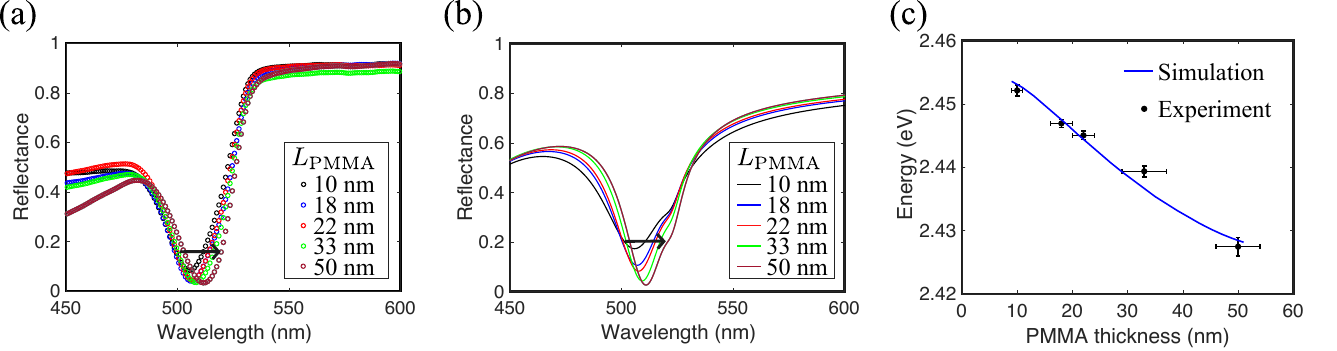}
\caption{(a) Experimental and (b) simulated reflectance spectra of our device architecture: PMMA/PEPI/PMMA/Ag/glass, when the PMMA thickness is increased from 10 nm to 50 nm. The arrow indicates redshift of the main dip as the PMMA thickness increases. Panel (b) is plotted in colour-to-colour correspondence to panel (a). (c) The main dip position of the reflectance as a function of PMMA thickness for both experiment (black dots) and simulation (blue curve). The energy errorbars are obtained from the fitting error of two-mode Gaussian functions on the experimental data, whereas the PMMA thickness errorbars correspond to the surface roughness of the PMMA layers.}
\label{FIG_pepipmma}
\end{figure*}

\subsection{Device architecture: PMMA/PEPI/PMMA/Ag/glass substrate}

As an additional test case, we study PEPI thin film in a typical device architecture, which involves an additional metallic electrode layer.
For simplicity, we consider a heterostructure design involving an end silver thin film, i.e., PMMA/PEPI/PMMA/Ag/Glass substrate, where we investigate the effect of the thickness variation of the Ag and PMMA layers to the excitonic mode splitting.
We will use the same parameters for the PEPI as those obtained from fitting in Table~\ref{tableFitting}, unless stated otherwise.

\vspace{0.3cm}
First, let us fix the thickness of both PMMA layers at $L_{\scriptsize \mbox{PMMA}}=10$~nm, and vary that of the silver layer $L_{\scriptsize \mbox{Ag}}$.
The results are shown in {\bf Figure~\ref{FIG_pepiID}}, for different PEPI layer thickness (units) in panels (a)-(c).
As expected, the reflectivity spectra of our device architecture are similar to those presented in Figure~\ref{FIG_pepiDT} for very thin silver layer, i.e., in the limit where $L_{\scriptsize \mbox{Ag}}$ goes to zero (black curves).
As the Ag layer thickness increases, the reflectance of the sample also increases, as the light transmitted through the sample is reflected back by the high-reflectivity metal layer.
An interesting observation can be seen in the intermediate PEPI layer thickness (Figure~\ref{FIG_pepiID}(b)).
As the Ag layer thickness increases from 10 nm (brown curves) to 100~nm (blue curves), it can be seen that the two excitonic modes present originally in the PEPI layer are becoming the two dips in the final heterostructure.
Figure~\ref{FIG_pepiID}(c) also shows that the second excitonic mode is more visible in the device architecture with thicker PEPI layer, as one would expect.
Also note that after a certain thickness of the silver layer $>100$~nm, our simulations show the reflectivity spectra converge close to the blue curves.

\vspace{0.3cm}
Now let us vary the thickness of both the PMMA layers $L_{\scriptsize \mbox{PMMA}}$, while the Ag layer is fixed at $L_{\scriptsize \mbox{Ag}}=80$~nm.
Our experimental data is presented in {\bf Figure~\ref{FIG_pepipmma}}(a), where the thickness of the top and bottom PMMA layers are increased from 10 nm to 50 nm.
Our simulations are plotted in Figure~\ref{FIG_pepipmma}(b) with color-to-color correspondence to panel (a). 
Note that we set the PEPI layer thickness corresponding to having 27 units.
Interestingly, we observe a redshift of the excitonic modes with increasing PMMA thickness, as indicated by the arrows.
The exciton mode position as a function of the PMMA layer thickess is summarized in Figure~\ref{FIG_pepipmma}(c).
Here, our model simulation shows an excellent agreement with the experimental data.
Note that in the simulations, we have used the exciton resonance energy of a single quantum well as $\hbar\omega_0$ = 2.385 eV, which is redshifted from the previous fitted value 2.387 eV.
\textcolor{black}{This shift is very small and it can be explained as due to the presence of the Ag layer, which can be treated with the method of image charges, giving an overall negative Coulomb energy between excitons and their mirror images (reduction in the potential energy of the excitons).}

\section{Conclusion}

In summary, we have shown that a PEPI perovskite thin film, owing to its naturally assembled quantum wells and strong coupling with light, can exhibit two pronounced excitonic modes.
By making use of the transfer matrix formalism, we derived an effective transfer matrix model for the PEPI layer, which enabled us to explain features observed in experiments.
Apart from fitting the experimental data, our model allowed us to gain insight to the origin of the mode splitting.
In particular, the emergence of the second mode is a result of light-mediated interactions between excitons in different quantum wells.
We showed both experimentally and with simulation that with thicker PEPI layers, the second mode becomes more apparent.
Furthermore, through PEPI/PMMA/PEPI structure design, we clarify the indirect coupling mechanism of excitons from the two PEPI layers via light.
We also showed that the thickness of the PMMA barrier provides tunability to the resulting excitonic modes.
Last, we presented a device architecture: PMMA/PEPI/PMMA/Ag/glass substrate.
We demonstrated both the enhancement and tunability of the excitonic modes with respect to the thickness of the silver and PMMA layers.

\vspace{0.3cm}
We have not only demonstrated the light-mediated exciton coupling in 2D perovskites, but also provided a realistic theoretical framework in which such a system can be described.
Our results shed new light on the possibility of multiple (more than two) excitonic modes, e.g., by having stacks of the PEPI/PMMA/PEPI structures.
Furthermore, one can place this new design in a microcavity, allowing further coupling of the excitonic modes with a strongly confined cavity light mode, extending the well-established strong exciton-photon coupling demonstrated in previously.\textsuperscript{\cite{strongexpho1,strongexpho2,strongexpho3,han2020transition}}


\begin{table*}[t]
\begin{center}
\caption{Summary of concentrations and spin-coating parameters for PEPI.
}
\label{table1}
\smallskip
\begin{tabular}{|c|c|c|c|}
\hline
Device configurations& \small{PEPI concentration}&\small{Spin-coating parameters}&$L_{\scriptsize \mbox{PEPI}}$\\
\hline
\hline
PEPI/glass&$0.125$~[{\scriptsize{M}}]& $4000$~[rpm], $30$~[s] &$47\pm5\mbox{ [nm]}$ \\
\hline
PEPI/glass&$0.125$~[{\scriptsize{M}}]& $5000$~[rpm], $30$~[s] &$37\pm5\mbox{ [nm]}$ \\
\hline
PEPI/glass&$0.125$~[{\scriptsize{M}}]& $6000$~[rpm], $30$~[s] &$27\pm2\mbox{ [nm]}$ \\
\hline
\small {PMMA/PEPI/PMMA/Ag/glass}&$0.125$~[{\scriptsize{M}}] & $4500$~[rpm], $30$~[s] &$40\pm3\mbox{ [nm]}$ \\
\hline

\end{tabular}
\end{center}
\end{table*}

\begin{table*}[t]
\begin{center}
\caption{Summary of PMMA concentrations, spin-coating parameters, and layer thickness.
}
\label{table2}
\smallskip
\begin{tabular}{|c|c|c|}
\hline
PMMA concentration& Spin-coating parameters&Thickness\\
\hline
\hline
$15$~[mg ml$^{-1}$]& $4000$~[rpm], $30$~[s] &$50\pm4$ [nm]\\
\hline
$10$~[mg ml$^{-1}$]& $4000$~[rpm], $30$~[s] &$33\pm4$ [nm]\\
\hline
$8.5$~[mg ml$^{-1}$]& $4000$~[rpm], $30$~[s] &$22\pm2$ [nm]\\
\hline
$7$~[mg ml$^{-1}$]& $4000$~[rpm], $30$~[s] &$18\pm2$ [nm]\\
\hline
$5$~[mg ml$^{-1}$]& $3500$~[rpm], $30$~[s] &$10\pm1$ [nm]\\
\hline

\end{tabular}
\end{center}
\end{table*}

\section{Experimental Section}
In order to fabricate highly crystallized perovskite thin films with full coverage and uniform surface morphology, the first essential step is to properly clean the substrates.
In this paper, glass substrates served as the base substrates.
They were cleaned in an ultrasonic machine for 15 min in each solvent in the following sequence: detergent, deionized water, acetone, and isopropanol.
After these wet cleaning steps, the substrates were further cleaned using a plasma cleaning machine for 10 min to remove remaining dusts and adjust the substrate surfaces from hydrophobic to hydrophilic.
PEPI precursors were prepared by dissolving phenethylammonium iodide ($\mbox{C}_6\mbox{H}_5\mbox{C}_2\mbox{H}_4\mbox{NH}_3\mbox{I}$ from \emph{Dyesol}) and lead iodide ($\mbox{PbI}_2$ from \emph{Acros Organics}) at a stoichiometric ratio (2:1) in Dimethylformamide (DMF from \emph{Sigma Aldrich}).
After mixing on a hotplate at $85^{\circ}$~C for 2 hours and filtering out any remaining impurities, PEPI precursors were spin-coated onto cleaned substrates in a nitrogen filled glovebox at certain spinning speed, as listed in {\bf Table~\ref{table1}}.
The corresponding thin-film thicknesses were calibrated by contact mode atomic force microscopy (AFM from \emph{Bruker Bioscope Resolve}).
A following annealing process at $100^{\circ}$~C for 10 min helped to remove solvents and crystallize into a stable perovskite phase.
\textcolor{black}{The information for crystallographic structure of PEPI thin films is shown in Appendix~\ref{Crystallographic}.}
The Polymethylmethacrylate (PMMA from \emph{Sigma Aldrich}) layers for device configuration of PMMA/PEPI/PMMA/Ag/glass were also fabricated by the spin-coating method.
The concentrations of PMMA solutions are summarized in {\bf Table~\ref{table2}}.
The silver layer was prepared by thermal evaporation under ultrahigh vacuum with an evaporation rate of $0.2$ \AA s$^{-1}$.

\begin{figure}[h]
\centering
\includegraphics[width=0.5\textwidth]{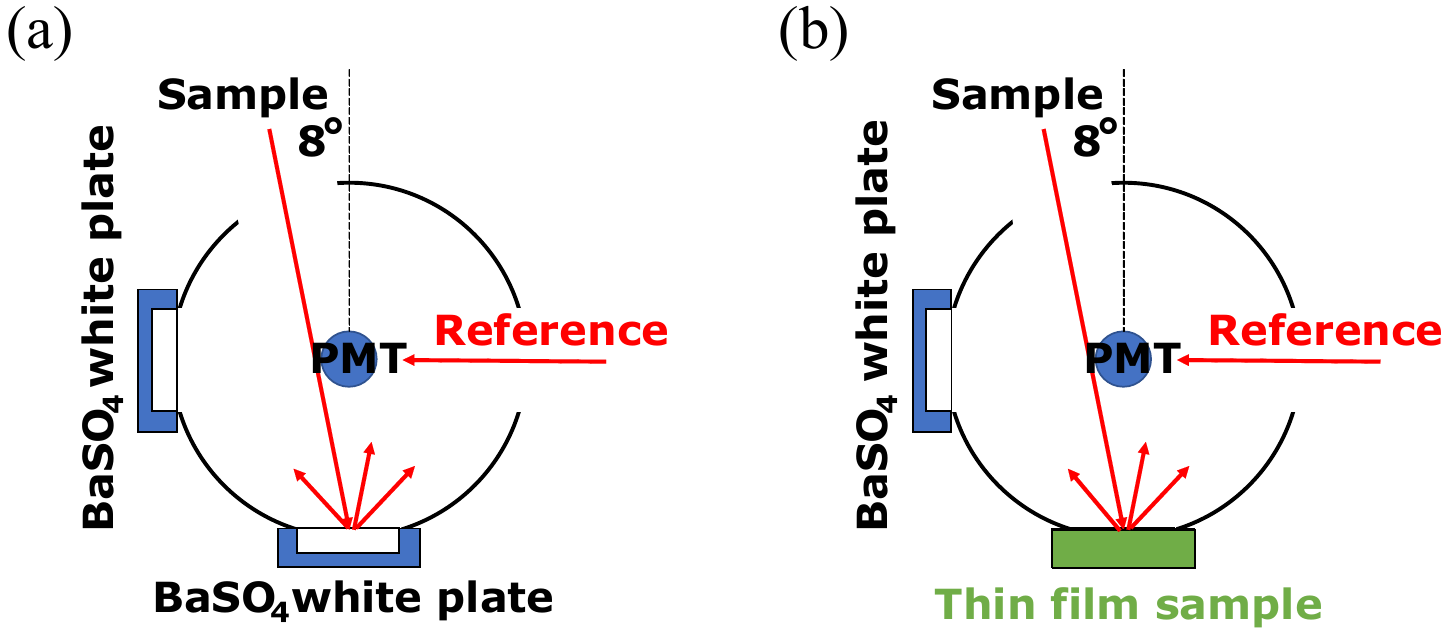}
\caption{Schematic procedures for steady-state total reflection measurements at room temperature.
}
\label{FIG_treflection}
\end{figure}

\vspace{0.3cm}
The steady-state total reflection spectra were measured using a UV-Vis-NIR spectrophotometer (\emph{Shimadzu UV-3600}) with an integrating sphere.
In the experimental setup, there were two beam paths (red arrows): one referred to as sample path was directed to a sample at an incidence angle of $8^{\circ}$, the other reference path was reared to a $\mbox{BaSO}_4$ white board at normal incidence, as indicated in {\bf Figure~\ref{FIG_treflection}}.
In this geometry, both diffuse reflection and specular reflection were collected by the central photomultiplier (PMT) detector.
Baseline correction (Figure~\ref{FIG_treflection}(a)) was the first step, where reflections were measured for both standard white boards.
The collected signals from the sample and reference beam paths are denoted by $I_{S1}$ and $I_{R1}$, respectively.
Then a sample was placed at the sample beam path (Figure~\ref{FIG_treflection}(b)) and the resulted signals were $I_{S2}$ and $I_{R2}$.
Thus, we obtained the total reflection of the sample as
\begin{equation}
R(\%)=\frac{I_{S2}/I_{R2}}{I_{S1}/I_{R1}} \times 100.
\end{equation}

\medskip

\section*{Acknowledgements}
T.K., Q.Z., K.D., and D.G. contributed equally to this work. 
T.K., D.G., T.C.H.L., and T.C.S. acknowledge support from the Singapore Ministry of Education under its AcRF Tier 2 grant MOE2017-T2-1-001. Q.Z. was supported by the Singapore Ministry of Education under its AcRF Tier 2 grant MOE2017-T2-2-002. K.D. was supported by the Singapore Ministry of Education under its AcRF Tier 2 grant MOE2019-T2-1-004. T.C.S. also acknowledges support from the Singapore National Research Foundation NRF-NRFI-2018-04 and AcRF Tier 2 grant MOE2019-T2-1-006.

\medskip

\newpage
\onecolumngrid
\appendix
\section{Reflectance}\label{M_reflectance}

In order to calculate the reflectance of a 2D heterostructure (with an effective transfer matrix ${\bf T}$), we assume that it is situated between a left material, in which the light is incident, and a right material (substrate), through which the light is transmitted and exits the heterostructure.
We will use the notation $n_{\scriptsize \mbox{L}}$ and $n_{\scriptsize \mbox{R}}$ for the refractive index of the material to the left and right of the heterostructure, respectively.

\vspace{0.3cm}
We set the amplitude of the electric field incident to the structure to one, the reflected and transmitted ones are noted by $r$ and $t$, respectively.
By taking the relation between the electric field $E_x$ and magnetic field $H_y$ into account, one can form the following transfer matrix equation:
\begin{equation}\label{EQ_Treflect}
{\bf T} \left[ \begin{array}{c} 1+r\\q_{\scriptsize \mbox{L}}(1-r) \end{array}\right]=\left[ \begin{array}{c} t\\q_{\scriptsize \mbox{R}}t \end{array}\right],
\end{equation}
where $q_{\scriptsize \mbox{L}}$ and $q_{\scriptsize \mbox{R}}$ are the inverse impedance of the left and right material.
The minus sign for the reflected magnetic field is due to the light propagating to the left.
One can now solve the reflectivity coefficient in Equation~(\ref{EQ_Treflect}) and obtain
\begin{equation}
r=\frac{-q_{\scriptsize \mbox{R}}T_{11}-q_{\scriptsize \mbox{L}}q_{\scriptsize \mbox{R}}T_{12}+T_{21}+q_{\scriptsize \mbox{L}}T_{22} }{q_{\scriptsize \mbox{R}}T_{11}-q_{\scriptsize \mbox{L}}q_{\scriptsize \mbox{R}}T_{12}-T_{21}+q_{\scriptsize \mbox{L}}T_{22}}.
\end{equation}
The reflectance of the 2D heterostructure is then given by $R=|r|^2$.
In the main text, we also utilise the transfer matrix method to show how the transverse electric field changes through the heterostructure, which can give insight for explaining the origin of the excitonic mode splitting.

\section{Refractive index: Glass, silver, and PMMA}\label{M_refractiveindex}

\begin{figure}[b]
\centering
\includegraphics[width=0.7\textwidth]{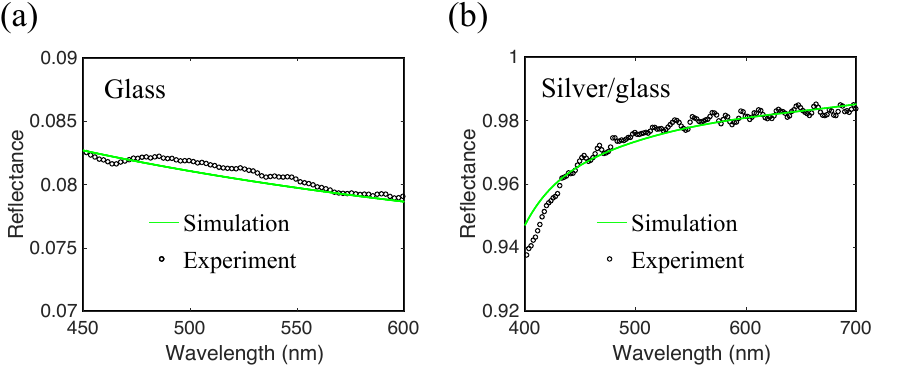}
\caption{Reflectivity spectra of normal layers.
Panels (a) and (b) are fittings of reflectance for the glass substrate and silver/glass substrate.
The black circles represent experimental data while the green curves are from simulations by utilising the transfer matrix method.
}
\label{FIG_nlayersR}
\end{figure}

Here we provide the refractive index dispersion model for glass, silver, and PMMA layers.
In the case of a  glass substrate we use a linear refractive index model, i.e., $n_{\scriptsize \mbox{G}}(\varepsilon)=\alpha+\beta \varepsilon$, where $\varepsilon$ is the energy of the probing light.
We found the coefficients $\alpha$ and $\beta$ by fitting the reflectance in our experiments.
In particular, we obtained
\begin{equation}
n_{\scriptsize \mbox{G}}=1.3900+0.0471\varepsilon,
\end{equation}
where $\varepsilon$ is in units of eV.
We plot the reflectivity fitting of the glass substrate in {\bf Figure~\ref{FIG_nlayersR}}(a).

\vspace{0.3cm}
Next, we model the refractive index for a silver layer by fitting the experimental data in previous work.\textsuperscript{\cite{nAg}}
The resulting refractive index dispersion, which includes the real and imaginary part of $n_{\scriptsize \mbox{Ag}}$, reads
\begin{eqnarray}
\mbox{Re}(n_{\scriptsize \mbox{Ag}})&=&0.0993-0.0201\varepsilon,\\
\mbox{Im}(n_{\scriptsize \mbox{Ag}})&=&11.2418-4.6356\varepsilon+0.5514\varepsilon^2.
\end{eqnarray}
We note that our model is based on fitting the experimental data in the range $400$-$700$~nm.
With this model we fit our experimental data for the reflectance of a silver layer on a glass substrate, see Figure~\ref{FIG_nlayersR}(b).
We used silver with $L_{\scriptsize \mbox{Ag}}\sim 80$~nm for the simulated green curve.
A thicker silver layer would result in the curve shifting upwards.

\begin{figure}[b]
\centering
\includegraphics[width=1\textwidth]{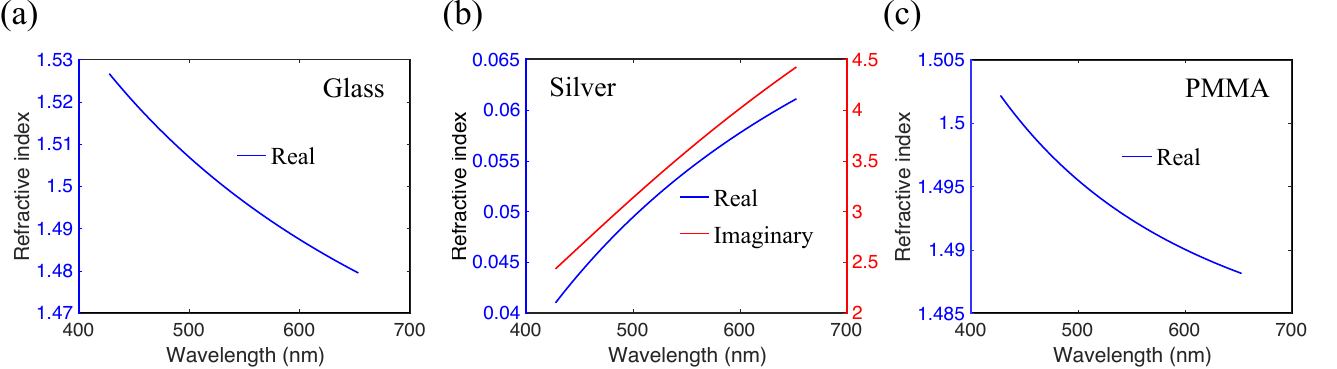}
\caption{Refractive index dispersion of normal layers.
The model refractive index is plotted for glass substrate (a), silver layer (b), and PMMA layer (c).
The blue curves represent the real component of the refractive index while the red curve indicates the imaginary part.
}
\label{FIG_nlayersRI}
\end{figure}

\vspace{0.3cm}
The experimental data in Ref.\textsuperscript{\cite{npmma}} was used to obtain the model of refractive index for PMMA layer:
\begin{equation}
n_{\scriptsize \mbox{PMMA}} = 1.4794-0.0015\varepsilon+0.0032\varepsilon^2,
\end{equation}
We used the acquired energy-dependent refractive index models and the thickness of silver layer as a reference, especially to fit the reflectivity spectra of heterostructures composed of more layers.
All the refractive index dispersions above are plotted in {\bf Figure~\ref{FIG_nlayersRI}}, where the solid blue curves (solid red curve) indicate the real (imaginary) component.

\section{Fitting: Percentage loss and background signal}\label{M_fitting}

\begin{figure}[b]
\centering
\includegraphics[width=0.7\textwidth]{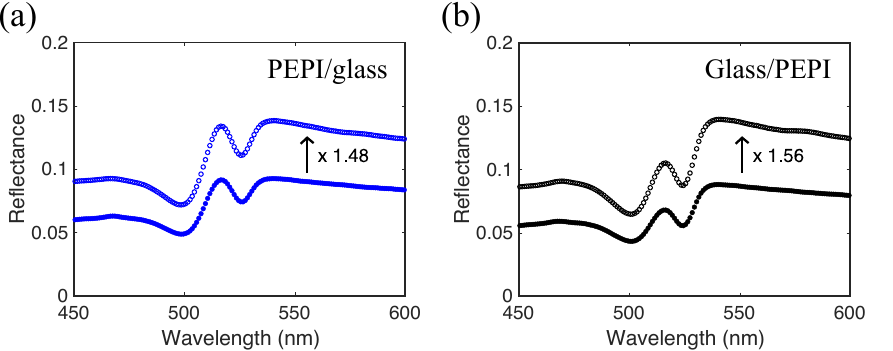}
\caption{Reflectivity spectra of PEPI/glass and glass/PEPI indicating percentage loss.
Panel (a) shows data where the light is incident on a PEPI layer situated on a glass substrate, while in panel (b) the probing light is incident from the opposite of that in panel (a).
The filled and unfilled circles in both panels represent data taken from the same sample but different spots. They are related by a constant, which is shown next to the arrow.
}
\label{FIG_pepiglass}
\end{figure}

We performed reflectivity spectra measurements for two cases where light is incident on the PEPI/glass and glass/PEPI, see {\bf Figure~\ref{FIG_pepiglass}} panels (a) and (b), respectively.
We note that the filled and unfilled circles represent data obtained differently (at different spots) from the same sample.
There is a fractional difference between the two sets of data for both the PEPI/glass and glass/PEPI cases.
This hints to the fractional loss of the reflected signals reaching the detector.

\begin{figure}[b]
\centering
\includegraphics[width=0.4\textwidth]{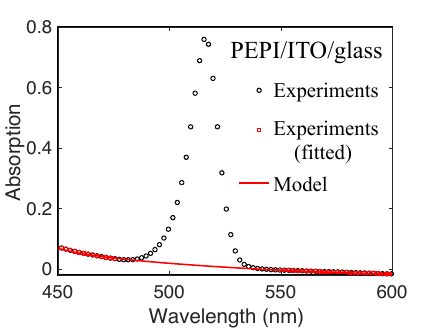}
\caption{Absorption spectrum of PEPI/ITO/glass substrate.
The black circles indicate the complete experimental data, while the red squares are the ones for fitting of the background signal.
We model the background signal as the solid red curve, which is a slowly varying function with respect to the wavelength of the probing light.
}
\label{FIG_background}
\end{figure}

\vspace{0.3cm}
Apart from the percentage loss, we also take into account a background signal, which is described below.
Based on the absorption spectrum of the structure: PEPI/ITO/glass substrate, there is a background signal that is weakly dependent on the energy of the probing light, see {\bf Figure~\ref{FIG_background}}.
We plotted the full absorption data as black circles, the selected ones for fitting (red squares), and the corresponding model for background signal (red curve).
More explicitly, we take the function
\begin{equation}
I_{\scriptsize \mbox{Bg}}=p_1\varepsilon^5 + p_2\varepsilon^4 +p_3\varepsilon^3 + p_4\varepsilon^2 +p_5\varepsilon + p_6,
\end{equation}
where $\varepsilon$ is the energy of the light in units of eV and the parameters obtained from fitting are 
\begin{equation}
(p_1,p_2,p_3,p_4,p_5,p_6)=(0.9304,-10.7425,49.5383,-113.9658, 130.8018,-59.9598).
\end{equation}

\vspace{0.3cm}
Hence, in our simulations, we plot the simulated reflectance as
\begin{equation}
R_{\scriptsize \mbox{plot}}=\eta R_{\scriptsize \mbox{ideal}}-\xi I_{\scriptsize \mbox{Bg}}+R_0,
\end{equation}
where $R_{\scriptsize \mbox{ideal}}$ is the ideal reflectance with all the reflected signals taken into account, $R_0$ is a constant background, $\eta$ is a fraction characterising the loss of the detected signal, and $\xi$ the strength of the background signal.
We note that important features in experiments such as the observed dips in reflectivity spectra, especially the centre (in frequency), are captured in $R_{\scriptsize \mbox{ideal}}$ and will not be affected by the background signal $I_{\scriptsize \mbox{Bg}}$ or $R_0$.

\section{\textcolor{black}{Crystallographic structure of PEPI thin films}}\label{Crystallographic}

\begin{figure}[b]
\centering
\includegraphics[width=0.45\textwidth]{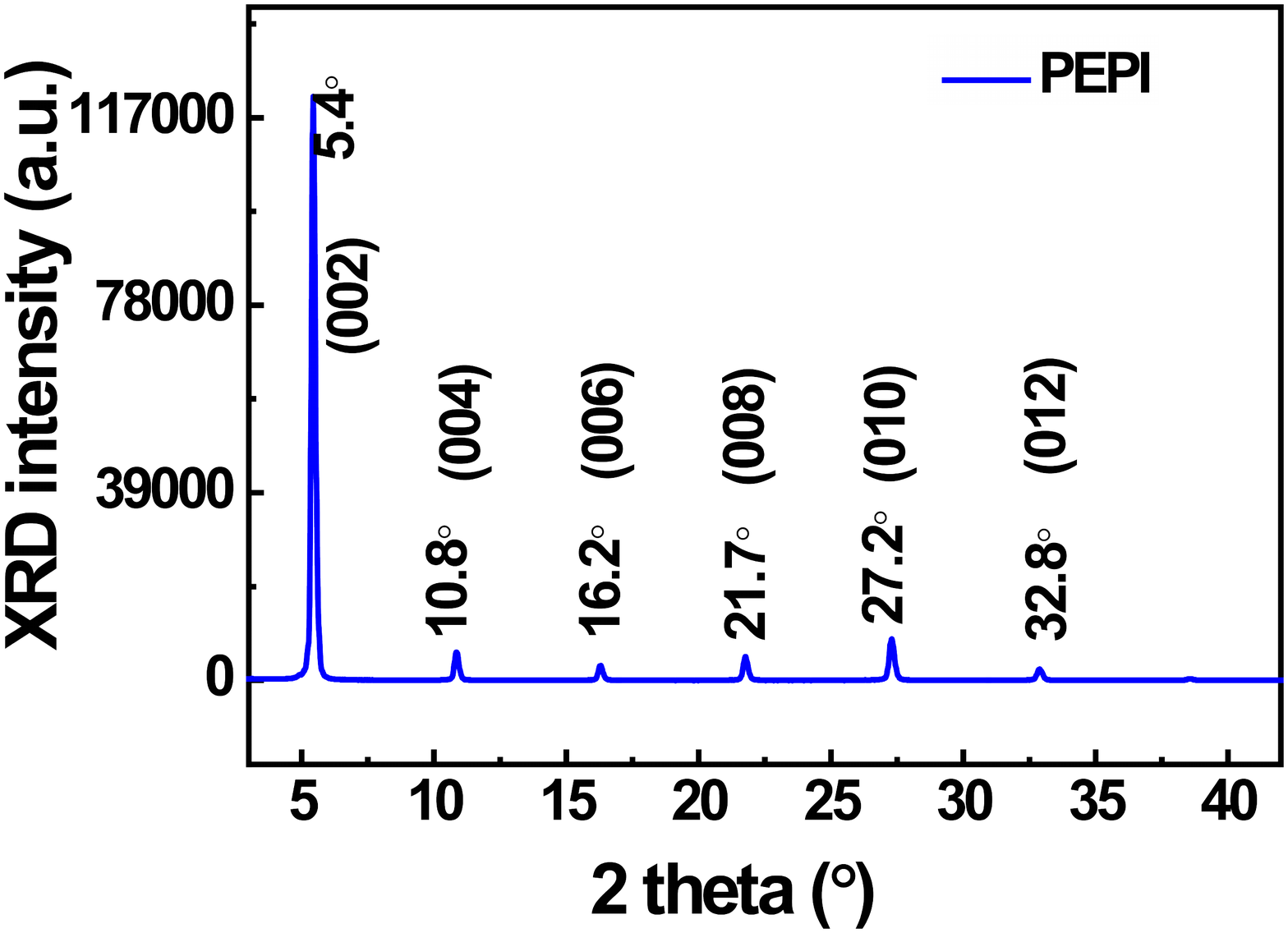}
\caption{\textcolor{black}{XRD spectrum of PEPI thin films. The strongest peak at $5.4^{\circ}$ corresponds to (002) plane. The existence of equally spaced diffraction peaks indicates that PEPI thin films are highly crystallized and the PEA layer planes are parallel to the substrate.}}
\label{FIG_XRD}
\end{figure}

\textcolor{black}{For the investigation of the crystallographic structure of PEPI thin films in our experiments, we performed X-ray diffraction (XRD) measurements at room temperature, see {\bf Figure~\ref{FIG_XRD}}, before we proceeded with optical measurements.
Equally spaced diffraction peaks at $5.4^{\circ},10.8^{\circ},16.2^{\circ},21.7^{\circ},27.2^{\circ},$ and $32.8^{\circ}$ indicate that PEPI thin films are highly crystallized and have layer-by-layer parallel structure. 
Since these XRD results provide only out-of-plane crystallization information of our samples, we know that each parallel layer of PEPI thin films are parallel to the substrate. 
Thus, these diffraction peaks correspond to $(00l)$ planes and the strongest peak at $5.4^{\circ}$ corresponds to $(002)$ planes.
We note that our XRD patterns for this archetypal 2D perovskites are consistent with previously reported work.\textsuperscript{\cite{perovstructure}}
According to Bragg's law, we obtained the interlayer spacing of two perovskite inorganic layers. For example, the interlayer distance of $(002)$ planes is $16.3\:\AA$.}

\clearpage

\end{document}